\documentclass[12pt]{article}
\newcommand{\be}{\begin{equation}}
\newcommand{\ee}{\end{equation}}
{

\begin{document}
\begin{center}
{\large \bf COHERENT STATES AND GEOMETRIC PHASES IN
CALOGERO-SUTHERLAND MODEL\footnote{Supported in part by grant No.
R02-2000-00040 and by grant No. R04-2000-00002 from the Korea
Science \& Engineering Foundation.} }
\end{center}
\begin{center}
Dae-Yup Song\\
{\small \it Department of Physics, Sunchon National University,
Suncheon 540-742, Korea\\
and}\\
JeongHyeong Park\\
{\small \it Department of Mathematics, Honam University, Gwangju
506-714, Korea}
\end{center}

\begin{abstract}
Exact coherent states in the Calogero-Sutherland models (of
time-dependent parameters) which describe identical harmonic
oscillators interacting through inverse-square potentials are
constructed, in terms of the classical solutions of a harmonic
oscillator. For quasi-periodic coherent states of the
time-periodic systems, geometric phases are evaluated. For the
$A_{N-1}$ Calogero-Sutherland model, the phase is calculated for a
general coherent state. The phases for other models are also
considered.
\end{abstract}

\newpage

\section{Introduction}
The Calogero-Sutherland (CS) models \cite{Cal,Suth} which describe
identical $N$-body harmonic oscillators interacting through
inverse square potentials in one-dimension have long been of great
interest. These models are closely related to the random matrix
model \cite{Mehta,GMW} and have been found relevant for the
descriptions of various physical phenomena \cite{SLA}. While the
generalization of the model on a circle without confining harmonic
potential (the Sutherland model) \cite{Suth2,Suth3} has been of
great interest \cite{SLA}, we also mention that generalization has
been considered in various directions \cite{HS,HR}. The underlying
algebraic structure of the models has been analyzed by the
exchange operator formalism \cite{Poly}, and the symmetric
polynomials in the wave functions of the models have been studied
through the quantum inverse scattering method \cite{UW1,UW2}.

The close relationship between the CS model and $N$-body
non-interacting harmonic oscillators has been noticed from the
inception of the model, and one of the prominent features of a
harmonic oscillator system is the existence of coherent states
\cite{Klauder,mmm}; A coherent state whose center of the
probability distribution function moves along a classical solution
can be given by applying a displacement operator to an eigenstate
of a constant Hamiltonian system, while a generalized coherent
state (squeezed state) whose width are described by the classical
solutions can be obtained through a unitary squeeze
transformation. For a CS model with harmonic potential, a
(squeezed-type) coherent state has been shown to exist
\cite{SuthCoh,BG}. Recently, it has been further shown that, the
coherent states of the CS models (of time-dependent parameters)
can be derived from the eigenstates of the model (of constant
parameters) through the (squeeze- and displacement-type) unitary
transformations \cite{ManyCoh}, as in the harmonic oscillator
\cite{Uni}. For the coherent states of the CS model of $N$-body
system, the center and width of the particle number density
function are described by the classical solutions of a harmonic
oscillator \cite{SuthCoh,ManyCoh}.

If a system is described by a periodic Hamiltonian and a wave
function is quasi-periodic (periodic up to a global phase factor)
under the time-evolution with (an integral multiple of) the
Hamiltonian's period,  a geometric phase, the geometric part of a
change in the phase of a wave function, can be defined
\cite{AA,Berry}. These geometric phases have attracted great
interest both theoretically and experimentally \cite{SW}. For the
CS models, the condition for the quasi-periodicity of a coherent
state is exactly same to that in the harmonic oscillator system
\cite{ManyCoh,Sgeo}, and steps forward the calculation of a
geometric phase has been done for a coherent state corresponding
to the ground state \cite{BCG,WBG}.

In this article, the coherent states and geometric phases of the
CS models with harmonic potential will be systematically studied.
It will be shown that, thanks to the orthogonality and recurrence
relations among the coherent states (or the eigenstates), the
geometric phases for the coherent states can be evaluated in terms
of the classical solutions of a harmonic oscillator, without
explicit knowledge of the normalization of the wave functions. For
the $A_N$ CS model where the interaction is written in terms of
the differences between the positions of two particles, both of
squeeze- and displacement-type unitary transformations can be
applied to give the coherent states, and the geometric phase will
be found as a sum of the contributions from the motion of the
center and from the squeezing of the coherent state. If the
symmetric polynomial in the coherent state is trivially 1, the
contribution from the squeezing is proportional to the energy
eigenvalue. The geometric phases for coherent states of the $B_N$
model and another model will also be considered.

In the next section, the known facts about $A_N$ model will be
recapitulated and the coherent states will be found. In section 3,
the geometric phases will be evaluated for a general coherent
state of the $A_N$ model. In section 4, the coherent states and
geometric phases of the $B_N$ and another model will be evaluated.
A summary will be given in the last section.

\section{The coherent states from eigenstates}
In this section we will consider the $A_{N-1}$ CS model.
The model of unit mass and angular frequency is described by the
Hamiltonian
 \begin{equation}
 H_A^s=\sum_{i=1}^N({p_i^2\over 2}+{x_i^2\over 2})
  +\sum_{i>j=1}^N {\hbar^2 \lambda(\lambda-1)\over(x_i-x_j)^2}.
 \end{equation}
By defining $y_N$ and $r$ as
 \begin{equation}
 y_N={1\over\sqrt{N}}(x_1+x_2+\cdots x_N),~~~
 r=\left(\sum_{i=1}^N x_i^2\right)^{1/2},
 \end{equation}
the (unnormalized) eigenstate can be written as \cite{GP}
 \begin{eqnarray}
 \phi_{m,n}(x_1,x_2,\cdots,x_N)&=&
     \exp\left(-{y_N^2\over 2\hbar}\right)H_m({y_N\over\sqrt{\hbar}}) \phi_n^C  \\
   &=& \left( \prod_{i>j=1}^N(x_i-x_j)^\lambda \right)
        \exp\left(-{r^2\over 2\hbar}\right) H_m({y_N\over\sqrt{\hbar}})
       L_n^b\left({1\over \hbar}(r^2-y_N^2)\right),~~~
 \end{eqnarray}
where
 \begin{equation}
b={1\over 2}(N-3)+{1\over 2}\lambda N(N-1).
 \end{equation}
In Eq. (4), $H_m$ and $L_n^b$ denote the Hermite and the
associated Laguerre polynomial, respectively, and the energy
eigenvalue of $\phi_{m,n}$ with positive integers $m,n$ is
 \begin{equation}
 E_{m,n}=\hbar(m+2n)+{\hbar \over 2}[N+\lambda N(N-1)].
 \end{equation}
By defining a new coordinate system $\{ y_i |i=1,2,\cdots N \}$
which satisfies the linear relation $\vec{y}=O\vec{x}$ with an
orthogonal matrix $O$, the $H_A^s$ can be written as
 \begin{equation}
 H_A^s=({p_{y_N}^2 \over 2}+ {y_N^2 \over 2}) + H_{C,s},
 \end{equation}
while $H_{C,s}$ depends only on $y_1,y_2,\cdots y_{N-1}$. In the
work by Calogero \cite{Cal}, $\phi_n^C(y_1,y_2,\cdots,y_{N-1})$
has shown to be an eigenstate of $H_{C,s}$ \cite{ManyCoh}.

The system described by the Hamiltonian
 \begin{equation}
H_A={1\over2}\sum_{i=1}^N\left({p_i^2\over
M(t)}+M(t)w^2(t){x_i^2}\right)
  +{\hbar^2 \lambda(\lambda-1)\over M(t)}\sum_{i>j=1}^N {1\over(x_i-x_j)^2},
 \end{equation}
is related to the system of $H_A^s$ through the unitary
transformations \cite{ManyCoh}. If the two linearly independent
solutions of the equation of motion
 \begin{equation} {d \over {dt}} (M \dot{{x}}) + M(t)w^2(t) {x} =0
 \end{equation}
are denoted as $u(t)$ and $v(t)$, and $\rho(t)$ is defined by
$\rho(t)=\sqrt{u^2+v^2}$,  with a time-constant $\Omega$ ($\equiv
M(t)[ \dot{v}(t)u(t) - \dot{u}(t)v(t)]$), the wave function
satisfying the Schr\"{o}dinger equation $i\hbar{\partial \over
\partial t}\psi_{m,n}=H_A\psi_{m,n}$ is given from $\phi_{m,n}^s$
as
\begin{eqnarray}
 \psi_{m,n}(t;\vec{x}) &=&\left({u-iv \over
   \rho(t)}\right)^{E_{m,n}/\hbar}
   U_fU_N\phi_{m,n}^s(x_1,x_2,\cdots,x_N)~~~~~~~~~~~~~~~~ \\
 &=&e^{i (N\delta_f+ M\dot{u}_f\sqrt{N}y_N)/ \hbar }\left({u-iv \over \rho} \right)^{m+2n}
   \left({u+iv \over \sqrt{\Omega}}\right)^{-(N+\lambda
   N(N-1))/2}  \cr
 &&\times
   \exp\left[-{1\over 2\hbar}\left({\Omega \over \rho^2}-iM {\dot{\rho} \over \rho}\right)
            (r^2-2\sqrt{N} y_N u_f +N u_f^2) \right] \cr
  &&\times H_m\left(\sqrt{\Omega\over \hbar} {y_N-\sqrt{N} u_f\over\rho}\right)
  L_n^b\left({\Omega \over \hbar\rho^2}(r^2-y_N^2)\right),
\end{eqnarray}
while the overdot denotes differentiation with respect to $t$. In
Eq. (11), $u_f$ is a linear combination of $u(t)$ and $v(t)$, and
$\delta_f$ is defined through the relation $\dot{\delta}_f= {1
\over 2}M(w^2 u_f^2 -\dot{u}_f^2).$ In Eq. (10), $U_N$ and $U_f$
are defined as
 \begin{eqnarray}
 U_N &=& ({\Omega \over \rho^2})^{N/4}
     \prod_{i=1}^N (\exp[{i \over 2 \hbar}M {\dot{\rho} \over \rho} x_i^2]
     \exp[-{1 \over 2}\ln({\rho^2 \over \Omega})x_i {\partial \over \partial
     x_i}]),\\
 U_f &=& e^{{i \over \hbar} N\delta_f}\prod_{i=1}^N (\exp[{i\over\hbar}M\dot{u}_fx_i]
     \exp[-{i\over \hbar}u_f p_i]).
 \end{eqnarray}
If $u,v$ are taken as $\cos t, \sin t$, respectively, the
$\psi_{m,n}$ describes a coherent state of the system of
Hamiltonian in Eq. (1) \cite{ManyCoh}.

A general eigenstate of the Hamiltonian in Eq. (1) is written as
\cite{Cal,UW1,UW2}
 \begin{equation}
 \phi_{m,n,\kappa}^s(x_1,x_2,\cdots,x_N)=\phi_{m,n}^s
 P_{\kappa}(x_1,x_2,\cdots,x_N),
 \end{equation}
where a partition, $\kappa$, is an integer vector
$(\kappa_1,\kappa_2,\cdots,\kappa_N)$. With the weight
$k=|\kappa|=\sum_{i=1}^N \kappa_i$, the energy eigenvalue of
$\phi_{m,n,\kappa}^s$ is given as $E_{m,n}+\hbar k~(\equiv
E_{m,n,k})$. It has been shown that $P_\kappa$ is a homogeneous,
translational invariant, symmetric polynomial of degree $k$. Even
though we do not know the explicit form of $P_{\kappa}$, from
these properties, the coherent states $\psi_{m,n,\kappa}$
corresponding to $\phi_{m,n,\kappa}$ can be written as
 \begin{equation} \psi_{m,n,\kappa}=\left({u+iv \over
 \Omega}\right)^{-k}\psi_{m,n} P_{\kappa}.
 \end{equation}

\section{The geometric phases for the $A_{N-1}$ model}
In this section, we will study the geometric phase of the
$A_{N-1}$ model through the definition given by Aharonov and
Anandan \cite{AA} which is a generalization of Berry's phase by
removing the restriction to adiabatic evolution \cite{SW}. Since
the geometric phase is defined for a periodic system, we will
assume the periodicity of $M(t)$ and $w^2(t)$ under the
time-evolution with period $\tau$
 \begin{equation} M(t+\tau)=M(t),~~~~w^2(t+\tau)=w^2(t).
 \end{equation}
For the quasi-periodicity of the coherent states given in the
previous section, as for the harmonic oscillator system, $\rho(t)$
and $u_f(t)$ must be periodic under the time-evolution of an
integral multiple of $\tau$. As analyzed by one of the authors
\cite{Sgeo}, for the cases that the two linearly independent
homogeneous solutions of Eq. (9) are bounded all over the time,
there exists a choice of classical solutions for periodic $\rho$
with the period $\tau'$, where $\tau'$ is $\tau$ or $2\tau$
depending on the model. If $u(t)$ and $v(t)$ are not periodic with
period $\tau'$, for the periodicity,  $u_f$ will be taken as $0$.
For a normalized quasi-periodic wave function $\psi$ of a
$\tilde{H}$ system with the global phase change $\chi$ such that
 \begin{equation}
 \psi(t+\tau';\vec{x})=e^{i\chi}\psi(t;\vec{x}),\end{equation}
the geometric phase $\gamma$ is given as \cite{AA}
 \begin{eqnarray}
 \gamma
 &=&\chi+{1\over\hbar}\int_0^{\tau'} <\psi|\tilde{H}|\psi> \cr
 &=&\chi+i\int_0^{\tau'}
         <\psi|{\partial \over \partial t}|\psi>.
 \end{eqnarray}

The global phase change $\chi_{m,n}$ of $\psi_{m,n}$ under the
time-evolution $\tau'$ can be found as
 \begin{equation} \chi_{m,n}(\tau')= -\left(m+2n+{N+\lambda N(N-1) \over
 2}\right)\int_0^{\tau'}{\Omega \over M\rho^2}dt.
 \end{equation}
In deriving Eq. (19), the fact that we only consider the periodic
$\rho$ and $u_f$ with period $\tau'$ has been used. The expression
of $H_A^s$ in Eq. (7) clearly shows that the quantum number $m$
comes from the motion of the center of mass which is described by
a free harmonic oscillator. This gives the orthogonality
 \begin{equation} <\psi_{m,n}\mid\psi_{m',n'}>=\tilde{C}_{m,n,n'}\delta_{m,m'},
 \end{equation}
with some (unknown) constants $\tilde{C}_{m,n,n'}$. If we only
consider a sector of $x_1<x_2<\cdots <x_N$, $H_A^s$ is a
Hermitian. With the relations in Eqs. (10,20), one can thus find
that the coherent states satisfy the orthogonality relation
 \begin{equation}
 <\psi_{m,n}\mid\psi_{m',n'}>=C_{m,n}\delta_{m,m'}\delta_{n,n'}.
 \end{equation}
Making use of this orthogonality and the recurrence relations in
the Hermite and Laguerre polynomials, one can find the following
relation
 \begin{eqnarray}
  {1\over i}{<\psi_{m,n}|{\partial \over\partial t}|\psi_{m,n}> \over
  <\psi_{m,n}|\psi_{m,n}> }
  &=&{d\over dt}\chi_{m,n}(t)
  +{Nu_f\over \hbar}{d\over dt}(M\dot{u}_f) +i(m+2n){\dot{\rho}\over \rho}  \cr
 && +i(m+2n+b+{3\over 2}) {\rho^2\over 2\Omega}
  {d\over dt}\left({\Omega\over \rho^2}-iM{\dot{\rho} \over
  \rho}\right).
\end{eqnarray}
Eq. (22) is obtained through a long algebra, while, for a simple
case, a similar procedure will be described in detail in the next
section. The geometric phase $\gamma_{m,n}$ for $\psi_{m,n}$,
therefore, reads as
 \begin{equation} \gamma_{m,n}={N\over \hbar}\int_0^{\tau'} M\dot{u}_f^2 dt +
  {E_{m,n}\over \hbar}\int_0^{\tau'}{M\dot{\rho}^2\over \Omega}dt.
 \end{equation}
For $\psi_{m,n,\kappa}$, the global phase change
$\chi_{m,n,\kappa}(\tau')$ is given as
 \begin{equation}
\chi_{m,n,\kappa}(\tau')=\chi_{m,n}(\tau') -k\int_0^{\tau'}{\Omega
\over M\rho^2}dt=-{E_{m,n,k}\over \hbar}\int_0^{\tau'}{\Omega
\over M\rho^2}dt
 \end{equation}
while the geometric phase $\gamma_{m,n,\kappa}$ under the
$\tau'$-evolution is same to that of $\psi_{m,n}$, as
 \begin{equation} \gamma_{m,n,\kappa}=\gamma_{m,n}.
 \end{equation}

Since the global phase change depends only on energy eigenvalue,
the geometric phase can be defined for a superposition of the
coherent states corresponding to eigenstates of the same energy
eigenvalue. Another eigenstate for the Hamiltonian $H_A^s$ can be
found as
 \begin{equation}
 \phi_n^s(x_1,x_2,\cdots,x_N)
   =  \prod_{i>j=1}^N(x_i-x_j)^\lambda
     \exp\left(-{r^2\over 2\hbar}\right)
       L_n^{b+{1\over 2}}\left({r^2\over \hbar}\right),
 \end{equation}
with the energy eigenvalues ${\hbar\over 2}[N+\lambda
N(N-1)]+2\hbar n=E_{0,n} $. From the identities among Hermite and
Laguerre polynomials, this eigenstate and the corresponding
coherent state $\psi_n$ for the system of $H_A$ are written in
terms of $\phi_{2l,n-l}^s$ and $\psi_{2l,n-l}^s$, as
 \begin{equation}
 \phi_n^s=\sum_{l=0}^n {(-)^l\over 2^{2l} l!} \phi_{2l,n-l}^s,~~~
 \psi_n=\sum_{l=0}^n {(-)^l\over 2^{2l} l!} \psi_{2l,n-l}^s.
 \end{equation}
The geometric phase of $\psi_n$ is given as $\gamma_{0,n}$.

\section{Other models}
For exposing the calculational procedure of geometric
phase through orthogonality and recurrence relations, we consider
a model described by the Hamiltonian
 \begin{eqnarray}
 H_w^s &=& \sum_{i=1}^N({p_i^2\over 2}+{x_i^2\over 2})
    +\sum_{i>j=1}^N {\hbar^2 \lambda(\lambda-1)\over(x_i-x_j)^2} \cr
 && +\sum_{i=1}^N{\alpha(\alpha-1)N(N-1)\over 2w_i^2}
    -\sum_{i>j=1}^N {2N\alpha(\alpha+N\lambda)\over w_iw_j},
 \end{eqnarray}
where $w_i=(\sum_{j=1}^Nx_j)-Nx_i.$ The (unnormalized) ground
state of this system is given as
 \begin{equation}
 \phi_0^w=\left(\prod_{i>j=1}^N(x_i-x_j)\right)^\lambda \left(\prod_{i=1}^N
 w_i \right)^\alpha \exp(-{r^2\over 2\hbar})
 \end{equation}
with energy eigenvalue $E_0^w=\hbar(\lambda{N(N-1)\over 2}+\alpha
N+{N\over 2}).$ For the case of $N=3$, from the fact that
$\sum_{i>j=1}^3 {1\over w_iw_j}=0$, one can find that this system
corresponds to the one which has long been known \cite{Wol,GKP}.
The case of $N=4$ has also been considered in the literature
\cite{GKP}. One of the excited states with the energy eigenvalue
$E_1=E_0+2\hbar$ for the system of $H_w^s$ is written as
 \begin{equation}
 \phi_1^w=(r^2-E_0)\phi_0^w.
 \end{equation}
For the system described by the Hamiltonian
 \begin{eqnarray}
 H_w&=&\sum_{i=1}^N{1\over 2}({p_i^2\over M(t)}+M(t)w^2(t)x_i^2)
  +{1\over M(t)}\sum_{i>j=1}^N {\hbar^2
  \lambda(\lambda-1)\over(x_i-x_j)^2}  \cr
  &&+{1\over M(t)}\left[\sum_{i=1}^N{\alpha(\alpha-1)N(N-1)\over 2w_i^2}
  -\sum_{i>j=1}^N {2N\alpha(\alpha+N\lambda)\over w_iw_j}\right],
 \end{eqnarray}
the coherent states are given as
\begin{eqnarray}
\psi_0^w &=& \left( {u+iv \over
\sqrt{\Omega}}\right)^{-E_0^w/\hbar}
\left(\prod_{i>j=1}^N(x_i-x_j)\right)^\lambda \left(\prod_{i=1}^N
 w_i \right)^\alpha \exp \left[-{r^2\over 2\hbar}\left({\Omega \over \rho^2}
 -iM{\dot{\rho} \over \rho}\right) \right], \\
\psi_1^w&=&\left({u-iv \over \rho}\right)^2\left( {\Omega \over
\rho^2}r^2-E_0^w\ \right)\psi_0^w.
\end{eqnarray}
Since the interaction can be written in terms of the differences
between positions of two particles, the displacement-type unitary
transformation can also be applied to give the coherent state; In
obtaining Eqs. (32-33), however, only squeeze-type transformation
has been used for simplicity. Furthermore, we only consider the
geometric phase for the coherent state $\psi_0^w$ which is
obtained from the ground state. From the fact that the Hamiltonian
is a Hermitian in a sector, the orthogonality relation is given as
 \begin{equation} <\psi_0^w|\psi_1^w>=0.
 \end{equation}
With the global phase change $\chi_0^w(\tau')$ which satisfies
 \begin{equation} \psi_0^w(t+\tau')=\exp[i\chi_0^w(\tau')]\psi_0^w(t),
 \end{equation}
one can find the relation
 \begin{equation}
 {1\over i}\dot{\psi}_0^w= \dot{\chi}_0^w(t)\psi_0^w
 +{i\over 2\hbar}{\rho^2\over \Omega}{d \over dt}({\Omega \over \rho^2} -iM{\dot{\rho}
 \over \rho})\left[ \left( {\rho\over u-iv} \right)^2\psi_1^w
 +E_0\psi_0^w \right].
 \end{equation}
Making use of the definition in Eq. (18) and the relation in Eq.
(34), one can find that the geometric phase of $\psi_0^w$ under
the $\tau'$-evolution is given as
 \begin{equation} \gamma_0^w=
  {E_0^w \over \hbar}\int_0^{\tau'}{M\dot{\rho}^2\over \Omega}dt.
 \end{equation}

 The $B_N$ CS model is described by the Hamiltonian
 \begin{equation}
 H_B^s =\sum_{i=1}^N({p_{y_i}^2 \over 2}+{y_i^2\over 2})
     +\sum_{i=1}^N{\hbar^2\alpha(\alpha-1) \over 2y_i^2}
    +\hbar^2\lambda(\lambda-1)\sum_{i>j=1}^N
          [{1 \over (y_i-y_j)^2}+{1 \over (y_i+y_j)^2}],
 \end{equation}
whose ground state is known as
 \begin{equation}
 \psi_0^B=\left(\prod_{i>j=1}^N (x_i^2-x_j^2)^\lambda \right)
          \left(\prod_{i=1}^N x_i^\alpha\exp(-{x_i^2\over
          2\hbar})\right)
 \end{equation}
with energy eigenvalue $E_0^B=\hbar({N\over 2}+\lambda
N(N-1)+\alpha N).$ By defining $\tilde{b}={N\over 2} +\lambda
N(N-1) +N\alpha-1$, an excited states of energy eigenvalue
$E_n^B=E_0^B+2\hbar n$  can be found as
 \begin{equation}
 \phi_n^B=\psi_0^B L_n^{\tilde{b}}({r^2\over \hbar}).
 \end{equation}
Since the interaction term can not be written in terms of
differences between the positions of two particles, only
squeeze-type unitary transformation can be applied to give
coherent states. For the system described by the Hamiltonian
 \begin{eqnarray}
 H_B^s &=& \sum_{i=1}^N{1\over 2}({p_{i}^2 \over M(t)}+
 M(t)w^2(t)x_i^2)  \cr
    && +\sum_{i=1}^N{\hbar^2\alpha(\alpha-1) \over 2 M(t)x_i^2}
    +{\hbar^2\lambda(\lambda-1)\over M(t)}\sum_{i>j=1}^N
          [{1 \over (x_i-x_j)^2}+{1 \over (x_i+x_j)^2}],
 \end{eqnarray}
the coherent state corresponding to $\phi_n^B$, therefore, reads
 \begin{eqnarray}
 \psi_n^B&=&
 \left( {u+iv \over \sqrt{\Omega}}\right)^{-E_0^B/\hbar} \left({u-iv \over\rho}\right)^2
 \left(\prod_{i>j=1}^N (x_i^2-x_j^2)^\lambda \right)
          \left(\prod_{i=1}^N x_i^\alpha\right)
 \cr
 &&\times
           \exp \left[-{r^2\over 2\hbar}\left({\Omega \over \rho^2}
 -iM{\dot{\rho} \over \rho}\right) \right]L_n^{\tilde{b}}
 \left({\Omega r^2 \over \hbar \rho^2}\right),
 \end{eqnarray}
and the geometric phase is given as
 \begin{equation} \gamma_n^B=
  {E_n^B \over \hbar}\int_0^{\tau'}{M\dot{\rho}^2\over \Omega}dt.
 \end{equation}

\section{Concluding remarks}
We have studied the coherent states and geometric phases in the CS
models. Making use of the orthogonality and recurrence relations
among the coherent states, the geometric phases for the coherent
states have been evaluated in terms of the classical solutions of
a harmonic oscillator.

The geometric phase can be written as a sum of the contributions
from the motion of the center of mass and from the motion of
squeezing. In the harmonic oscillator system, the contribution
from the squeezing is proportional to energy eigenvalue of the
corresponding eigenstate \cite{Sgeo}, while, if the symmetric
polynomial in the wave function is not trivially 1, this is not
true in the CS models. We believe that the method developed in
this paper could be applied for the calculations of geometric
phases in the other CS models with confining harmonic potential.

\end{document}